\documentclass[11pt]{openjournal}
\usepackage{tabularx}
\usepackage{graphicx}
\usepackage{cancel}
\usepackage{multirow}
\usepackage{amssymb}
\usepackage{bm}

\usepackage{amsmath}
\usepackage{comment}

\usepackage{color}
\usepackage[pdftex, colorlinks=true, linkcolor=black, urlcolor=black,  citecolor=blue]{hyperref}
\usepackage{subfigure} 
\usepackage{float}
\begin{document}
\title{An independent assessment of significance of annual modulation in  COSINE-100 data}
\author{Aditi Krishak$^1$}
\email{aditi16@iiserb.ac.in}
\author{Shantanu Desai$^2$}
\email{shntn05@gmail.com}
\affiliation{$^{1}$ Department of Physics, Indian Institute of Science Education and Research Bhopal, Madhya Pradesh 462066, India}
\address{$^2$Dept. of Physics, Indian Institute of Technology Hyderabad,
Kandi, Telangana 502285, India}

\begin{abstract}
\noindent
We perform an independent search for annual modulation caused by dark matter-induced scatterings  in the recently released COSINE-100 data. We test the hypothesis  that  the  data contains a  sinusoidal modulation against the null hypothesis that the data consists of only background. We compare the significance using frequentist, information theoretic techniques (such as  AIC and BIC), and also using the Bayesian model comparison technique. The information theory-based tests mildly prefer a constant background over a sinusoidal signal with the same period as that found by the DAMA collaboration. The Bayesian test  however strongly prefers a background model.  This is the first proof of principles demonstration of application of Bayesian and information theory based techniques to COSINE-100 data to assess the significance of annual modulation.
\end{abstract}

\maketitle

\section{Introduction}
Although about  25\%  of the universe's matter density consists of cold dark matter~\citep{Planck18}, we have no clue about the mass of the dark matter particle or its non-gravitational couplings~\citep{Kamionkowski}. The most theoretically favored  and widely studied cold matter candidate is the Weakly Interacting Massive Particle (or WIMP)~\citep{Weinberg}. A large number of experiments have been taking data for more than 30 years to look for direct signatures of WIMP-nucleon interactions~ in underground laboratory-based experiments~\citep{Schumann}. Among these, only the DAMA/LIBRA experiment has detected an annual  modulation,   having all the right characteristics  of been induced by WIMPs in our galaxy~\citep{Freese88}, with a statistical significance of about 12$\sigma$~\citep{Dama18}.
However, the WIMP parameter space inferred from the DAMA/LIBRA results is ruled out by many other direct detection experiments. Although many attempts (for eg.~\citealt{Zupan,Catena, Gelmini,Williams,Tomar}, and references therein) have been made to reconcile the
results of DAMA with the null results of  other  experiments using non-standard particle physics or astrophysics assumptions, the jury is still out on whether any of them can satisfactorily reconcile with the latest results from all the direct detection experiments.
The only possible resolution out of this conundrum could be that, no other direct detection experiment with null results  used the same target material as DAMA, viz. thallium-doped NaI. The COSINE-100 experiment~\citep{Cosine} is one of the first experiments, whose detector is designed to be a replica of the DAMA target,  and  hence can   confirm or refute their annual modulation claims in a model-independent fashion. Many other experiments, designed to do a similar  test of the DAMA annual modulation such as  DM-Ice17~\citep{dmice}, KIMS~\citep{Kims}, SABRE~\citep{Sabre}, and ANAIS-112~\citep{Anais} are also about to start taking data  and  the ANAIS experiment has released preliminary results.

In a recent work~\citep{Krishak}, we did an independent assessment of the DAMA/LIBRA annual modulation claims from their most recent data release,  using three disparate model comparison techniques: frequentist~\citep{Desai16b}, Bayesian~\citep{Trotta,Weller}, and information theoretic techniques~\citep{Liddle,Liddle07}. The Bayesian and information theoretical techniques are widely used for model comparison in Astrophysics and Cosmology, but rarely used   in direct dark matter detection experiments. In this work, we apply the same techniques to the recently released data from the COSINE-100 experiment~\citep{Cosine}.

The outline of this paper is as follows. A brief summary of the COSINE-100 results can be found in Sect.~\ref{sec:cosine100}. Our own re-analysis is described in Sect.~\ref{sec:analysis}. We conclude in Sect.~\ref{sec:conclusions}. We do not provide any details of the theory behind the different model comparison techniques used herein, which can be found in \citet{Krishak} and references therein. Our analysis codes and results can be found on a  {\tt github} link, whose url is provided in Sect.~\ref{sec:conclusions}.

\section{Recap of Cosine-100 results}
\label{sec:cosine100}
We provide a brief recap of the main results in \citet{Cosine} (CS100 hereafter), wherein more details can be found. The COSINE-100 experiment is located   at the Yangyang underground
laboratory in  South Korea under more than 700~m of rock overburden. The experiment consists of eight  NaI crystals (labeled C1 to C8) doped with thallium and was designed to mimic the DAMA/LIBRA setup as closely as possible. Out of these, data from three crystals was omitted due to various systematics, as discussed in CS100. Data taking commenced in October 2016 and the results released in CS100 correspond to a total exposure of 97.7 kg years. The count rates for the five crystals used for the analysis can be found in Fig. 3 of CS100. The event rates were fit to the following functional form:
\begin{equation}
R = C+ p_0\exp (-\frac{\ln2 \cdot t}{p_1}) + A \cos (w (t-t_0)) .     
\label{eq:1}
\end{equation}

The first two terms in Eq.~\ref{eq:1}, consisting of the constant and exponential decay are used for parameterizing the background rates and the  last cosine term  is a potential signature of annual modulation caused by dark matter interactions. The data from all the crystals were   simultaneously fit to the same values of the cosine function parameters, but separately for $C$, $p_0$ and $p_1$ using $\chi^2$ minimization. Their results are consistent within $1\sigma$ with  both the null hypothesis of no oscillation as well as with the DAMA/LIBRA annual modulation best-fit values in the 2-6 keV range. The best fit parameters for different scenarios (phase fixed as well as floating) can found in Table 1 of CS100.

\section{Our Analysis}
\label{sec:analysis}
For our analysis, we obtained the data points and the errors associated with them from the COSINE-100 collaboration. The data consists of event rates for crystals 2, 3, 4, 6, and 7 in the 2-6 keV energy bin in 15-day intervals.
We first fit only the background rates (first two terms in Eq.~\ref{eq:1}) to the data and determine  the best-fit values for $C$, $p_0$ and $p_1$; this model  is assumed to be  our null hypothesis $H_0$, i.e.,
\begin{equation}
H_0(t) = C+ p_0\exp (-\frac{\ln2 \cdot t}{p_1}).
\label{eq:2}
\end{equation}
We then determine the  estimates of the best-fit parameters of the sinusoidal modulation  in Eq.~\ref{eq:1}, and this is considered as the hypothesis to be tested, viz. $H_1$. These two models are compared using frequentist, information theory (AIC and BIC), and Bayesian model comparison techniques. More details about these techniques have been recently reviewed  in \citet{Krishak} and references therein, and we skip these details for brevity.

\subsection{Parameter Estimation} \label{ssec:1}

Parameter estimation for the models under consideration is the first step towards model comparison analysis. 
The data points consist of experimental errors in the event rates ($\sigma_i$). For the model with only the background signal, we find the best-fit values of the parameters using $\chi^2$ minimization for each crystal separately.
The $\chi^2$ functional between the data ($y_i$) and the model function ($H(t)$) is given by:
\begin{equation}
\chi^2= \sum_{i=1}^N\left( \frac{y_i-H(t)}{\sigma_{i}}\right)^2 ,
\label{eq:chisq}
\end{equation}
where $y_i$ denotes the COSINE-100 event rate in time bin $i$ for each crystal, and $H(t)$ is the model is defined in Eq.~\ref{eq:2}. All the background parameters are kept free, with a positive constraint (lower bound of $10^{-10}$) on all of them; and the best-fit values obtained for each crystal by $\chi^2$ minimization are summarized in Table~\ref{table:1}.

\begin{figure*}
\centering
\includegraphics[width = 0.9\textwidth]{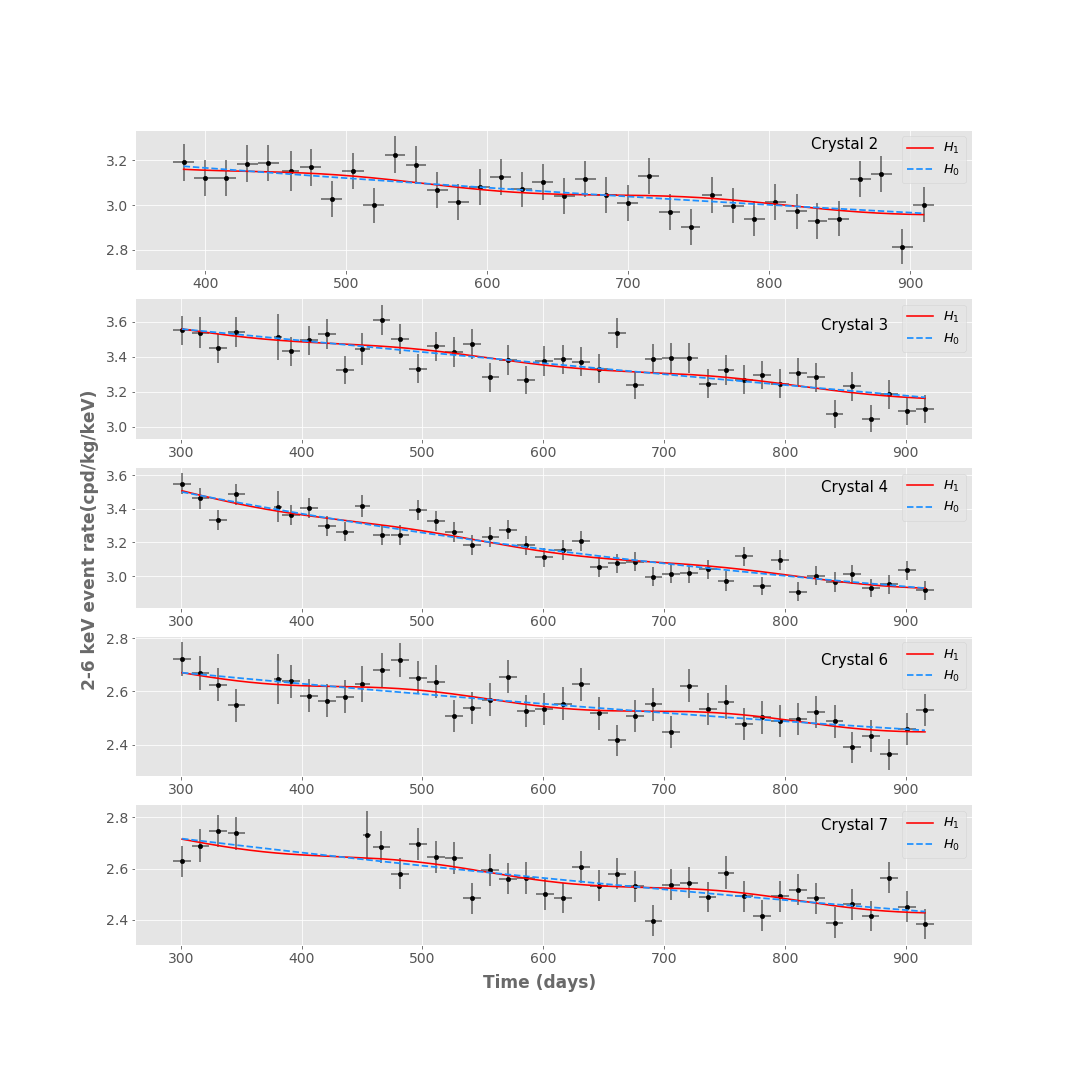}
\caption{The CS100 data points (in black) for all the five crystals are overlaid with the fits calculated for both the hypotheses, sinusoidal modulation with all parameters free $H_1(t)$  (Eq.~\ref{eq:1}, shown in red), and background-only model $H_0(t)$ (Eq.~\ref{eq:2}, shown in blue). The best-fit value of $\omega$ in this case is equal to 0.024 radians/day or period of  about 257 days.\smallskip}
\label{fig2}
\end{figure*}

For the model with a sinusoidal modulation (where  H(t) is defined in Eq.~\ref{eq:1})), the $\chi^2$ minimization is done concurrently for all the crystals by using the same values of $A$, $\omega$, and $t_0$ for  all the crystals, while the background parameters can be different for each crystal.  We first do a $\chi^2$ minimization keeping all the 18 parameters free. In this case, since the time bin width we have used is equal to 15 days, we are not sensitive to periods less than 15 days. Therefore, while doing the fits,  we have also used a lower bound on the period, equal to  15 days. The optimization  is done using the {\tt SLSQP}~\citep{slsqt} constrained optimization algorithm 
as implemented in {\tt scipy Python} module, keeping a positive lower bound on all background parameters, as well as on the amplitude and frequency.
The best fits obtained are listed in Table~\ref{table:2}. These fits along with the data can be found in Fig.~\ref{fig2}.  The best-fit value for  $\omega$  is about 0.024 radian/day (corresponding to a period of about 257 days).


For testing the DAMA annual modulation claim, we also  carry out optimization of this model by keeping the modulation frequency fixed at the DAMA obtained value of  0.0172 radians/day (or period fixed at 365.25 days). We then redo the  $\chi^2$ minimization with 17 free parameters (which is one less than before), with lower bound on all background parameters and on the amplitude. The best-fit for this optimization can be found in Table~\ref{table:4}. 


 The values obtained by us for the background parameters for each of the  crystals in both the cases, i.e. keeping all parameters free (Table~\ref{table:2}) and then keeping $\omega$ fixed (Table~\ref{table:4}), differ significantly from those obtained by the COSINE-100 collaboration\footnote{Although these values are not displayed in CS100, these were obtained by private communication with the authors of CS100. }, which is due to the  degeneracy between the background parameters $C$, $p_0$ and $p_1$ in Eq.~\ref{eq:1}.
 
 We now present model  comparison results using both these fits.

\begin{table}
    \centering
\begin{tabular}{cccc}
    \hline 
    {} &
    \textbf{$C$}{(cpd/keV/kg)} &
    \textbf{$p_0$}{(cpd/keV/kg)}&
    \textbf{$p_1$}{(days)}\\ 
    \hline 
    
    Crystal 2 & 2.48 & 0.90 & 995.59 \\
    Crystal 3 & 0.00 & 3.77 & 3675.20 \\
    Crystal 4 & 2.52 & 1.50 & 485.11 \\
    Crystal 6 & 2.05 & 0.77 & 1000.20 \\
    Crystal 7 & 1.90 & 1.01 & 1000.05 \\
    \hline
\end{tabular}
\caption{Best-fit values for each of the five crystal for the background-only model, consisting of the first two terms in Eq.~\ref{eq:1}.}
\label{table:1}
\end{table}

\begin{table}
    \centering
\begin{tabular}{ccccccc}
    \hline
    {} &
    \textbf{$C$}{(cpd/keV/kg)}&
    \textbf{$p_0$}{(cpd/keV/kg)}&
    \textbf{$p_1$}{(days)}&
    \textbf{$A$}{(cpd/keV/kg)}&
    \textbf{$\omega$}{(radians/day)}&
    \textbf{$t_0$}{(days)}\\ 
    \hline
    Crystal 2 & 2.0 &  0.88 & 994.55 &&& \\
    Crystal 3 & 0.01 & 3.76 & 3675.01 &&&\\
    Crystal 4 & 2.61 & 1.47 & 421.46 & 0.013 & 0.024 & 235.32\\
    Crystal 6 & 2.05 & 0.77 & 1008.91 &&&\\
    Crystal 7 & 1.91 & 0.99 & 993.22 &&&\\
    \hline
\end{tabular}
\caption{Best-fit values for the background plus cosine model in Eq.~\ref{eq:1} for each of the five crystals, when all the parameters are kept free. }
\label{table:2}
\end{table}

\begin{table}
    \centering
\begin{tabular}{cccccc}
    \hline
    {} &
    \textbf{$C$}{(cpd/keV/kg)}&
    \textbf{$p_0$}{(cpd/keV/kg)}&
    \textbf{$p_1$}{(days)}&
    \textbf{$A$}{(cpd/keV/kg)}&
    \textbf{$t_0$}{(days)}\\ 
    \hline
    Crystal 2 & 2.48 & 0.90 & 995.79 && \\
    Crystal 3 & 0.00 & 3.77 & 3674.63&&\\
    Crystal 4 & 2.51 & 1.52 & 486.23 & 0.009 & 133.2 \\
    Crystal 6 & 2.03 & 0.79 & 1001.07 &&\\
    Crystal 7 & 1.88 & 1.04 & 1000.21 &&\\
    \hline
\end{tabular}
\caption{Best-fit values for the background plus cosine model in Eq.~\ref{eq:1} for each of the five crystals, keeping the parameter $\omega$ fixed at $0.0172$ radians/day corresponding to a period of one year.}
\label{table:4}
\end{table}

\begin{figure*}
\centering
\includegraphics[width = 0.9\textwidth]{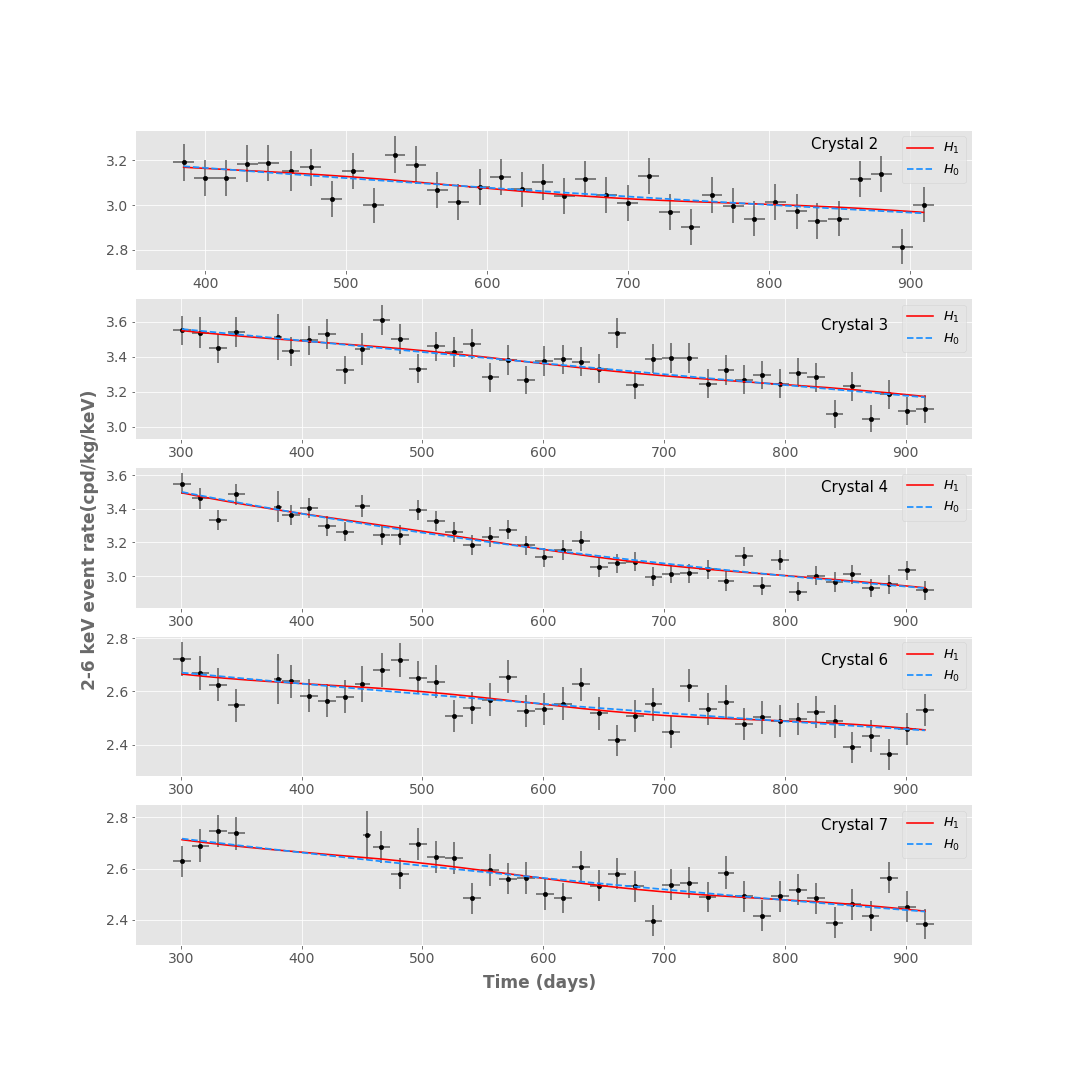}
\caption{The CS100 data points (in black) with the fits calculated for sinusoidal modulation $H_1(t)$ (Eq.~\ref{eq:1}, in red) with period fixed at 365.25 days, and background-only model $H_0(t)$ (Eq.~\ref{eq:2}, in blue). As we can see, by eye it is hard to distinguish between the two models. The data was obtained from the COSINE-100 collaboration (private communication).\smallskip}
\label{fig1}
\end{figure*}

\subsection{Model Comparison}
\subsubsection{Frequentist Model Comparison}
We carry out  frequentist model comparison by first calculating the $\chi^2$ values using Eq.~\ref{eq:chisq} with the best-fit parameters for each model, summed over all the data points for all five crystals. Then, using the best-fit $\chi^2$ and degrees of freedom, the goodness of fit for each model can be calculated from the  $\chi^2$ p.d.f.  The model with the greater value of $\chi^2$ p.d.f. would be considered as the favored model.

Making use of the fact that the two models are nested, we use Wilk's theorem \citep{Wilks} to quantify the $p$-value of the cosine model as compared to the background model. For our example, the difference in $\chi^2$ between the two models satisfies a $\chi^2$ distribution with degrees of freedom equal to three. From the cumulative distribution of $\chi^2$, we obtain the $p$-value from the $\chi^2$ c.d.f. The corresponding significance or $Z$-score is calculated using the prescription in \citet{Cowan11}. High $p$-value and low $Z$-score indicate weak evidence against the null hypothesis. The $\chi^2$ values per degree of freedom and the model  likelihood given by the $\chi^2$ p.d.f.  calculated for each model for both the cases ($\omega$ varying and $\omega$ fixed) can be found in Tables~\ref{table:5} and~\ref{table:6} respectively along with the $p$-value and $Z$-score. As we can see, for both the cases (with $\omega$ varying and $\omega$ fixed), the difference in $\chi^2$   between the two hypothesis is negligible. The significance of annual modulation with the same period as DAMA data is negligible (less than 1$\sigma$.)

\subsubsection{Information Criteria}
The Akaike Information Criterion value (AIC) is given by~\citep{Liddle07}:
\begin{equation}
\rm{AIC} = \chi^2_{min} + 2p,
\label{eq:AIC}
\end{equation}

The Bayesian Information Criterion is given by~\citep{Liddle07}:
\begin{equation}
\rm{BIC} = \chi^2_{min} + p \ln N, 
\label{eq:BIC}
\end{equation}
where $p$ is the number of free parameters, $\chi^2_{min}$ is the minimum $\chi^2$ value  and $N$ is the total number of data points. The model with the smaller value of AIC and BIC is preferred. We then calculate the difference in AIC and BIC values between the $H_1$ and $H_0$ hypothesis, and evaluate the significance using the qualitative strength of evidence rules given in ~\citet{Shi}. For the case with all modulation parameters free, the $\Delta$AIC and $\Delta$BIC values are tabulated in Table~\ref{table:5}. We see that the modulation hypothesis has much smaller values for AIC and BIC, when $\omega$ is a free parameter. However, according to the strength of evidence rules,  $|\Delta$AIC$|$ and $|\Delta$BIC$|$ have to be greater than 10 for the model with the smaller value to be decisively favored compared to the other. When $\omega$ is a free parameter, this criterion is not satisfied for AIC, so the better model cannot be decisively favored, whereas BIC test favors the null hypothesis.
For the case of  modulation period fixed, the $\Delta$AIC and $\Delta$BIC values are tabulated in Table~\ref{table:6}. We get smaller values for both AIC and BIC for the null hypothesis of background-only model in this case. However, since the  $\Delta$AIC and  $\Delta$BIC values are less than 10, they  also do not decisively favor any one model over the other. According to strength of evidence rules~\citep{Shi}, BIC shows strong evidence for the background only hypothesis.
Therefore, the background only hypothesis is mildy preferred using the information theory based tests.
\vspace{5mm}
\subsubsection{Bayesian Model Comparison}
We carry out a Bayesian model comparison by calculating the  Bayes factor  $B_{21}$ for the $M_2$ model in comparison to the $M_1$ hypothesis. Here, we consider the null hypothesis ($H_0$) to be $M_1$ and the cosine model ($H_1$) to be $M_2$.  The Bayes factor is given by~\citep{Trotta}:
\begin{equation}
B_{21} = \frac{P(D | M_2)}{P(D|M_1)} ,
\end{equation}
 where $P(D | M_2)$ and $P(D|M_1)$ are the marginal likelihood or Bayesian evidence for $M_2$ and $M_1$ respectively given data $D$. Similar to the previous model comparison tests, we calculated the Bayes factor for two cases: when $\omega$ is fixed at the DAMA best-fit value as well as when $\omega$ is a free parameter. Unlike the previous three tests, this statistic does not use the best-fit value of the parameters.
 
 We first calculate the Bayesian evidence for both $H_0$ and $H_1$ using the multi-threaded {\tt Dynesty} package~\citep{dynesty} in {\tt Python}, which uses the Dynamical Nested Sampling algorithm for calculating the Bayesian evidence~\citep{multinest,Mukherjee}. The likelihood function ($P(D|M,\theta)$) for the combined data from all the five crystals, given the model and a set of parameters, is assumed  to be a Gaussian:
\begin{equation}
P(D|M,\theta)= \prod_{j=1}^5   \left(\prod_{i=1}^N \frac{1}{\sigma_{total}\sqrt{2\pi}} \exp\left[-\frac{1}{2}\left(\frac  {y_i-H(t)}{\sigma_{total}}\right)^2\right]\right),
\label{eq:5}
\end{equation}
where $H(t)$ is in the form described in Eq.~\ref{eq:1}, $N$ is the total number of data points in each crystal, and the outer product is over the five crystals used for the analysis. We then multiply the likelihood by priors for all the background parameters as well as  for $A$ and $t_0$ and $\omega$ (when it is kept as a free parameter).

We choose uniform priors between $[0,400]$, $[0,400]$, and $[0,30000]$ for the background parameters $C$, $p_0$ and $p_1$ respectively. These bounds are conservative and cover a huge swath of parameter space for the background parameters. Outside these bounds, the calculation of Bayesian evidence also does not always converge. For the signal parameters of the sinusoid, we use uniform priors 
for  $A$ between $[0,400]$, and $[0,360]$ for $t_0$. When $\omega$ is kept as a free parameter, we choose a uniform prior between $[0.0104,0.428]$ rad/day, which corresponds to periods between 15 (bin size) and 600 days (maximum duration of the dataset).

The values of the Bayes factor for both the fits can be found in Tables~\ref{table:5} and~\ref{table:6}. We find that in both the cases the Bayes factor is less than 1, indicating that the background model is favored over the cosine-based fit. 
We use the Jeffrey's scale~\citep{Trotta} for a qualitative interpretation of the Bayes factor. Since $|\ln (B_{21})| > 10$, in both the cases, this  provides a strong evidence in favor of the null hypothesis.

\begin{table}
    \centering
    \begin{tabular}{|c|cc|}
    \hline
        {} & {$H_0$} & {$H_1$}\\
        \hline
        \textbf{Frequentist} & {} & {}\\
         {$\chi^2$}/DOF & 174.2/180 & 170.9/177 \\
         {$\chi^2$ p.d.f.} & 0.0207 & 0.0208\\
         {$p$-value}    & \multicolumn{2}{c|}{0.34}\\
         {significance} & \multicolumn{2}{c|}{0.42 $\sigma$} \\
         \hline
         \textbf{AIC} & 204.2 & 206.8 \\
         {$\Delta$ AIC} & \multicolumn{2}{c|}{2.6} \\
         \hline
         \textbf{BIC}  &253.3 & 265.8\\
         {$\Delta$ BIC}  & \multicolumn{2}{c|}{12.5}\\
         \hline
         $\mathbf{\ln (B_{21})}$
         & \multicolumn{2}{c|}{-16}\\
         \hline 
 \end{tabular}
\caption{Summary of model comparison results using frequentist, Bayesian and information theoretic criterion for $H_0$ (background only) and $H_1$ (model of background+cosine modulation with all parameters free). According to the frequentist model comparison test, the $\chi^2$ p.d.f (which represents the likelihood of the model given the data) for $H_1$ hypothesis is   almost the same as  that for $H_0$ hypothesis. The $H_0$  hypothesis has smaller values for BIC and AIC. However, only the difference in BIC  is greater  than 10, and hence  BIC   decisively favors the null hypothesis according to the  strength of evidence rules. The Bayesian model comparison test  based on the calculation of Bayes factor, however provides  ``strong evidence'' for the null hypothesis using the Jeffreys scale.}
 \vspace{5mm}
 \label{table:5}
\end{table}


\begin{table}

    \centering
    \begin{tabular}{|c|cc|}
    \hline
        {} & {$H_0$} & {$H_1$}\\
        \hline
        \textbf{Frequentist} & {} & {}\\
         {$\chi^2$}/DOF & 174.2/180 & 172.6/178 \\
         {$\chi^2$ p.d.f.} &  0.0207 & 0.0209 \\
         {$p$-value}    & \multicolumn{2}{c|}{0.44}\\
         {significance} & \multicolumn{2}{c|}{0.14 $\sigma$} \\
         \hline
         \textbf{AIC} & 204.2 & 206.6 \\
         {$\Delta$ AIC} & \multicolumn{2}{c|}{2.4} \\
         \hline
         \textbf{BIC}  & 253.3 & 262.2\\
         {$\Delta$ BIC}  & \multicolumn{2}{c|}{8.9}\\
         \hline
         $\mathbf{\ln (B_{21})}$
         & \multicolumn{2}{c|}{-7}\\
         \hline 
 \end{tabular}
\caption{Summary of model comparison results using frequentist, information theoretic, and Bayesian   criterion for $H_0$ (background only) and $H_1$ (background+cosine modulation with $\omega$ fixed.) According to the frequentist model comparison test, the $\chi^2$ p.d.f (which represents the likelihood of the model given the data) for $H_1$ hypothesis is   almost the same as  that for $H_0$ hypothesis.  The null hypothesis has a smaller value of AIC and BIC, but the difference does not cross the threshold of 10 for any one model to be decisively favored. The BIC test shows strong evidence for $H_0$ hypothesis. However, the Bayesian model comparison test  based on the calculation of Bayes factor,  provides very strong evidence for the null hypothesis using the Jeffreys scale.}
 \vspace{5mm}
 \label{table:6}
\end{table}

\section{Conclusions}
\label{sec:conclusions}
Recently, the COSINE-100 experiment, designed to test the DAMA/LIBRA annual modulation hypothesis, released their first results from  their search for annual modulation, induced from  dark matter scatterings, using  1.7 years of data, with a total exposure of 97.7 kg years~\citep{Cosine}. They find that the data in the 2-6 keV energy interval is consistent with both the null hypothesis of no modulation as well as with the DAMA estimate of  amplitude and phase at 68.3\% c.l.

In this work, we apply (similar to the analysis done in \citet{Krishak} for the DAMA/LIBRA data) three independent model comparison techniques, viz. frequentist, Bayesian and information theory-based, to test the compatibility of the data with annual modulation over a background-only hypothesis.

For the signal hypothesis, we did two different sets of fits. For one fit, we kept the period (or angular frequency) same as the DAMA best-fit value of one year or 0.0172 radians/day. For the other fit, the period was also kept as a free parameter.

Our results using all the three techniques are tabulated in Tables~\ref{table:5} and ~\ref{table:6} respectively. When $\omega$ is a free parameter, the BIC test decisively prefer the background-only model, whereas the significance of both the hypotheses is almost the same for the AIC and frequentist  test, and so its not possible to favor any one model from these tests.

When the period is fixed to the DAMA best-fit value of 1 year, we find  that the BIC test strongly  favors the background only hypothesis, but the difference in BIC  does not cross the threshold of 10 for it to be decisively favored. With the frequentist test, the difference is negligible.  With more data it remains to be seen if the significance increases with the frequentist and information theory based tests.

On the other hand, when we do the model comparison using Bayesian method of computing the Bayes factor, we find that the data strongly favors a constant background over a cosine fit (irrespective of whether $\omega$ is free or not).

This is the first proof of principle application of Bayesian and information theory based model comparison techniques to  the COSINE-100 data and is complementary to the statistical tests done in the COSINE-100 results paper. To promote transparency in data analysis, we have made our analysis codes and data publicly available, which can  be found at \url{https://github.com/aditikrishak/COSINE100_analysis}.

\section{Acknowledgements}
Aditi Krishak has been supported by DST-INSPIRE fellowship. We are grateful to Jay Hyun Jo  and the COSINE-100 collaboration for providing us the raw data used in Fig~\ref{fig1} and sharing with us their best-fit values of the background parameters. We acknowledge useful correspondence with Josh Speagle regarding the {\tt dynesty} algorithm.   We are also grateful to the anonymous referee for constructive feedback on the manuscript.

\bibliography{main.bib}
\end{document}